\documentclass[10pt,prb]{revtex4}
\usepackage{epsfig}
\usepackage{amsmath}

\begin{document}

\title{Non-adiabatic transition probability with a moving $\delta$ potential coupling}
\author{ Diwaker and Aniruddha Chakraborty \\
School of Basic Sciences, Indian Institute of Technology Mandi, Mandi, Himachal-Pradesh 175001, India.}

\begin{abstract}
The present work focuses on the calculation of a non-adiabatic transition probability between two states which may or may not cross with each other and are coupled to each other by a moving $\delta$ function potential. Here, the time dependent Schrodinger equation is converted to time independent one by using a scaling factor which is function of time. This time independent Schrodinger equation is then considered for two potentials coupled by a moving $\delta$ potential and an expression for non-adiabatic transition probability has been derived.  
\end{abstract}

\maketitle

\section{Introduction:}
\noindent
In time dependent Schrodinger equation which generally obey time dependent processes there are two types of time dependencies. One is where the time dependence is assumed for the adiabatic parameter while other is a quantum mechanical problem in a certain time dependent external field. As an example to explain the applicability of the former type is atomic collision processes where nuclear motion have time dependence and laser technology is of latter type. Literature survey reveals various anlytical models where the time dependent schrodinger equations are solved exactly. Some of the examples may include time dependent harmonic oscillator\cite{diw1, diw2, diw3}, infinite potential well with a moving boundary \cite{diw4, diw5, diw6} and various other examples \cite{diw7}. Two types of time-dependent problems which are commonly investigated includes the simple scattering problem where the potential has a periodic dependence on time \cite{diw8}and amounts to a steady state solution while the other type includes the delta functional potential where strength varies with time which is solved using Floquet formalism \cite{diw9}. Other time dependent problems includes the scattering problems involving delta potentials whose solutions are simple plane waves\cite{diw7}. There also exists time dependent problems which are solved using diffusive solution to the Schroedinger wave equation. Different kind of approaches are adopted by different authors to solve the time dependent problems which includes a path integral of Feynman type and other is Laplace Transforms. The main disadvantage of the former method is slow convergence which is rectified by introducing a rapidly converging scheme \cite{diw10}. The majority of the work involving the time dependent potentials is solved by using delta or a rectangular barrier. Intrinsically time dependent processes such as quantum mechanical processes in a certain time dependent external field becomes very important these days because of remarkable progress of Laser technology which means that Laser intensity and frequency can be now designed as a function of time. Our earlier work in this area was dedicated towards time independent coupling \cite{diw11, diw12, diw13, diw14,diw15,diw16,diw17}. In the present work we have studied the non adiabatic transition probability between two states which are coupled to each other by a moving $\delta$ function potential. We use a similar methodology \cite{diw18} by which we transfer our time dependent Schrodinger equation to independent one coupled by a moving $\delta$ potential by time dependent scaling factor and hence derived an analytical expression to calculate the non adiabatic transition probability thereby describing an important step to deal with the non adiabatic transition probability when we have moving $\delta$ potential coupling. We have organized our paper in the following manner. In section II we have discussed the methodology of conversion of time dependent Schrodinger equation to time independent one by using the time dependent scaling factor. Applicability of this method to the two state model coupled by moving $\delta$ potential method and derivation of non adiabatic transition probability is covered in section III while section IV concludes the paper.

\section{Transformation of time-dependent Schrodinger equation into time-independent Schrodinger equation using a time dependent scaling factor}
We start with a similar methodology applied  to our two state model as discussed by Lee \cite{diw18} and also by Berry and Klein \cite{diw19} where the problem of meta stability of a particle of mass $\mu$ trapped in a potential U(x,t) with a scaling form namely U(x,t) = $\overline{U}(\frac{x}{R(t)})\frac{1}{R^2(t)}$ with R(t) as a time dependent scaling factor is used to transform the time-dependent Schrodinger equation into  independent one. We consider time dependent Schrodinger equation for a two state model  as
\begin{equation}
 i\hbar \frac{\partial}{\partial t}\left[
\begin{array}{c}
\phi _1(x,t) \\
\phi _2(x,t)
\end{array}
\right] = \left[
\begin{array}{cc}
H_{11} & U_{12} \\
U_{21} & H_{22}
\end{array}
\right]\left[
\begin{array}{c}
\phi _1(x,t) \\
\phi _2(x,t)
\end{array}
\right].
\end{equation} 
where
\begin{eqnarray}
H_{11} = -\frac{\hbar^2}{2\mu}+V_{1}(x)\\
H_{22} = -\frac{\hbar^2}{2\mu}+V_{2}(x) \\
U_{12} = U_{21} = \frac{\overline{U}}{R^2(t)}\left(\frac{x}{R(t)}\right)
\end{eqnarray}
where the scaling factor $R(t)$ is assumed to be a linear function of time $R(t) = R_{0}+vt,$ with v = constant. Further in the rescaled frame with rescaled coordinate $\overline{x}$ = $\frac{x}{R(t)}$, equation $(1)$ can be written as
\begin{eqnarray}
i\hbar\frac{\partial}{\partial t}\phi_{1}(x,t) = \left[-\frac{\hbar^2}{2\mu}\frac{\partial^2}{\partial x^2}+V_{1}(x)\right]\phi_{1}(x,t)+\left[\frac{\overline{U}}{R^2(t)}\left(\frac{x}{R(t)}\right) \right]\phi_{2}(x,t)\nonumber \\
i\hbar\frac{\partial}{\partial t}\phi_{2}(x,t) = \left[-\frac{\hbar^2}{2\mu}\frac{\partial^2}{\partial x^2}+V_{2}(x)\right]\phi_{2}(x,t)+\left[\frac{\overline{U}}{R^2(t)}\left(\frac{x}{R(t)}\right) \right]\phi_{1}(x,t)\nonumber \\
\end{eqnarray}
In the two state model our lower potential $V_{1}(x) = 0$ hence above equations considerably reduces to 
\begin{eqnarray}
i\hbar\frac{\partial}{\partial t}\phi_{1}(x,t) = \left[-\frac{\hbar^2}{2\mu}\frac{\partial^2}{\partial x^2}\right]\phi_{1}(x,t)+\left[\frac{\overline{U}}{R^2(t)}\left(\frac{x}{R(t)}\right) \right]\phi_{2}(x,t)\nonumber \\
i\hbar\frac{\partial}{\partial t}\phi_{2}(x,t) = \left[-\frac{\hbar^2}{2\mu}\frac{\partial^2}{\partial x^2}+V_{2}(x)\right]\phi_{2}(x,t)+\left[\frac{\overline{U}}{R^2(t)}\left(\frac{x}{R(t)}\right) \right]\phi_{1}(x,t)\nonumber \\
\end{eqnarray}
In the rescaled frame as mentioned above equation $(6)$ can be further written as
\begin{eqnarray}
i\hbar\frac{\partial}{\partial t}\phi_{1}(\overline{x},t) = \left[-\frac{\hbar^2}{2\mu R^2}\frac{\partial^2}{\partial x^2}\right]\phi_{1}(x,t)+\left[ i\hbar \frac{v}{R}\overline{x}\frac{\partial}{\partial \overline{x}}+\overline{U}(\overline{x})\right]\phi_{2}(\overline{x},t)\nonumber \\
i\hbar\frac{\partial}{\partial t}\phi_{2}(\overline{x},t) = \left[-\frac{\hbar^2}{2\mu R^2}\frac{\partial^2}{\partial x^2}+V_{2}(x,t)\right]\phi_{2}(x,t)+\left[ i\hbar \frac{v}{R}\overline{x}\frac{\partial}{\partial \overline{x}}+\overline{U}(\overline{x})\right]\phi_{1}(\overline{x},t)\nonumber \\
\end{eqnarray}
Further the equation given below is obtained by making the substitution given in equation $(9)$ and $(10)$
\begin{eqnarray}
i\hbar\frac{\partial}{\partial t}\phi_{1}(\overline{x},\tau) = -\frac{\hbar^2}{2m}\frac{\partial^2}{\partial \overline{x}^2}\phi_{1}(\overline{x}, \tau)+\overline{U}_{12}(\overline{x})\phi_{2}(\overline{x},\tau)\nonumber \\
i\hbar\frac{\partial}{\partial t}\phi_{2}(\overline{x},\tau) = \left(-\frac{\hbar^2}{2m}\frac{\partial^2}{\partial \overline{x}^2}+V_{2}(x)\right)\phi_{2}(\overline{x}, \tau)+\overline{U}_{12}(\overline{x})\phi_{1}(\overline{x},\tau)\nonumber \\
\end{eqnarray}
with 
\begin{eqnarray}
\phi_{1}(\overline{x},t) = \frac{1}{\sqrt{R(t)}}e^{\left(\frac{i\mu}{2\hbar}\right)Rv\overline{x}^2}\psi_{1}(\overline{x},t)\nonumber \\
\phi_{2}(\overline{x},t) = \frac{1}{\sqrt{R(t)}}e^{\left(\frac{i\mu}{2\hbar}\right)Rv\overline{x}^2}\psi_{2}(\overline{x},t)\nonumber \\
\end{eqnarray}
and introducing a new variable 
\begin{equation}
\tau = \int_{0}^{t}\frac{ds}{R^2(s)} = \frac{t}{R_{0}R(t)}
\end{equation}
equation $(8)$ resembles the Schrodinger equation with a stationary state potential which can be solved by the separation of variables technique. i.e.
\begin{equation}
\psi(\overline{x},\tau) = \psi(\overline{x})e^{-\left(\frac{i}{\hbar}\right)\overline{E}\tau}
\end{equation}
where $\Psi(\overline{x})$ satisfy the eigen value equation
\begin{equation}
\left[-\frac{\hbar^2}{2m}\frac{d^2}{dx^2}+\overline{U}(\overline{x})\right]\psi_{k}(\overline{x}) = \overline{E}_{k}\psi_{k}(\overline{x})
\end{equation}
Above equation can be exactly solvable in the rescaled frame where the exact wave function in the rescaled frame is given by
\begin{equation}
\phi_{k}(x,t) = \frac{1}{\sqrt{R(t)}}e^{\left(\frac{i\mu}{2\hbar}\right)\left(\frac{v}{R}\right)x^2}e^{-\left(\frac{i}{\hbar}\right)\left(\frac{1}{R_{o}R}\right)\overline{E}t}\psi_{k}\left(\frac{x}{R}\right)
\end{equation}
where the set of solutions $(13)$ is complete and orthonormal.
\begin{equation}
\langle \phi_{k}(x,t)|\phi_{l}(x,t)\rangle = \langle \phi_{k}(\overline{x})|\phi_{l}(\overline{x})\rangle = \delta_{kl}
\end{equation}
furthermore if an initial state $\phi(x,0)$ is expressible in the basis $\{\psi_{k}\}$ as
\begin{equation}
\phi(x,0) = \sum_{k}c_{k}\phi_{k}(x,0), c_{k} = \langle\psi_{k}(x,0)|\psi(x,0)\rangle
\end{equation}
then at a later time t the state is given as
\begin{equation}
\psi(x,t) = \sum_{k}c_{k}\phi_{k}(x,t)
\end{equation}
\section{Non adiabatic transition probability for two state coupled by a moving $\delta$ potential coupling}
Using the methodology discussed in the preceding section we will consider the case of crossing of two diabatic potentials coupled by moving $\delta$ potential. The time dependent Schrodinger equation in this case can be written as
\begin{equation}
 i\hbar \frac{\partial}{\partial t}\left[
\begin{array}{c}
\phi _1(x,t) \\
\phi _2(x,t)
\end{array}
\right] = \left[
\begin{array}{cc}
H_{11} & U_{12} \\
U_{21} & H_{22}
\end{array}
\right]\left[
\begin{array}{c}
\phi _1(x,t) \\
\phi _2(x,t)
\end{array}
\right].
\end{equation} 
where
\begin{eqnarray}
H_{11} = -\frac{\hbar^2}{2\mu}+V_{1}(x)\\
H_{22} = -\frac{\hbar^2}{2\mu}+V_{2}(x) \\
U_{12} = U_{21} = \frac{\overline{U}_{o}}{R(t)}\delta(x-a(t)) = \frac{\overline{U}_{o}}{R^2(t)}\delta\left(\frac{x}{R(t)}-\overline{a}\right)
\end{eqnarray} 
where a(t) = $\overline{a}\;$R(t)$>0$ gives the position of the $\delta$ potential. In our case we have V$_{1}$ = 0 and potential V$_{2}$ is incorporated into energy E giving a new energy E$^{'}$ i.e. $E^{'} = E-V_{2}$ which is further equals to E for simplicity. In the rescaled frame as discussed in the last section our coupling is transformed as $U_{12} = U_{21} = \overline{U}_{o}\delta(\overline{x}-\overline{a})$ hence, our time dependent equations will be converted into time independent one and in the matrix notation can be written as
\begin{equation}\left[
\begin{array}{cc}
H_{11}&\overline{U}_{o}\delta(\overline{x}-\overline{a})\\
\overline{U}_{o}\delta(\overline{x}-\overline{a})&H_{22}
\end{array}\right]\left[\begin{array}{c}
\phi_{1}(\overline{x})\\
\phi_{2}(\overline{x})\\
\end{array}\right] = \overline{E}\left[\begin{array}{c}\phi_{1}(\overline{x})\\
\phi_{2}(\overline{x})\\\end{array}\right]
\end{equation}
The above equation is equivalent to the following 
\begin{eqnarray}
H_{11}\phi_{1}(\overline{x})+\overline{U}_{o}\delta(\overline{x}-\overline{a})\phi_{2}(\overline{x}) = \overline{E}\phi_{1}(\overline{x})\\
H_{11}\phi_{2}(\overline{x})+\overline{U}_{o}\delta(\overline{x}-\overline{a})\phi_{1}(\overline{x}) = \overline{E}\phi_{2}(\overline{x})
\end{eqnarray}
equation $(23)$ can be written as
\begin{equation}
\phi_{2}(\overline{x}) = \left(E-H_{22}\right)^{-1}\overline{U}_{o}\delta(\overline{x}-\overline{a})\phi_{1}(\overline{x})
\end{equation}
using equation $(24)$ into equation $(22)$ we get\cite{diw20} 
\begin{equation}
H_{11}\phi_{1}(\overline{x})+\overline{U}_{o}\delta(\overline{x}-\overline{a})\left(E-H_{22}\right)^{-1}\overline{U}_{o}\delta(\overline{x}-\overline{a})\phi_{1}(\overline{x}) = \overline{E}\phi_{1}(\overline{x})
\end{equation}
which will gave us finally 
\begin{equation}
H_{11}\phi_{1}(\overline{x})+\overline{U}_{o}^2 G_{2}^{o}(0,0,E)\delta(\overline{x}-\overline{a})\phi_{1}(\overline{x}) = \overline{E}\phi_{1}(\overline{x})
\end{equation}
to simplify further the above equation can be written as
\begin{equation}
H_{11}\phi_{1}(\overline{x})+\overline{V}_{o}\delta(\overline{x}-\overline{a})\phi_{1}(\overline{x}) = \overline{E}\phi_{1}(\overline{x})
\end{equation}
equation$(27)$ is achieved by making the following substitutions $ G_{2}^{o}(0,0,E) = \langle \overline{x}|\frac{1}{(E-H_{22}}|\overline{a}\rangle $and $\overline{V}_{o} = \overline{U}_{o}^2 G_{2}^{o}(0,0,E)$.
Above equation is a time independent Schrodinger equation in $\phi_{1}(\overline{x})$ whose general solutions can be written as 
\begin{gather*}
\begin{cases}
\phi_{1}(\overline{x}) = sin(\overline{k}\overline{x}),\; \overline{x} < \overline{a} \\
\phi_{1}(\overline{x}) = A \;cos(\overline{k}\overline{x}+\theta), \;\overline{a} < \overline{x}
\end{cases}
\end{gather*}
where $\overline{k} = \sqrt{\frac{2 \mu \overline{E}}{\hbar}}$, A is real constant and $\theta$ is a phase angle. The wavfunction and its derivative must satisfy the following boundary conditions at $\overline{x} = \overline{a}$ i.e.
\begin{eqnarray}
\phi_{1}(\overline{x} = \overline{a}^{+}) = \phi_{1}(\overline{x} = \overline{a}^{-}) \nonumber \\
and \nonumber \\
\frac{d \phi_{1}(\overline{x})}{d\overline{x}}|_{(\overline{x} = \overline{a}^{+})}-\frac{d \phi_{1}(\overline{x})}{d\overline{x}}|_{(\overline{x} = \overline{a}^{-})} = \frac{2\mu}{\hbar^2}\overline{V}_{o}\phi_{1}(\overline{x} = \overline{a})
\end{eqnarray}
from the above relations the coefficient A as a function of k can be determined as 
\begin{equation}
A^2(\overline{k}) = sin^2(\overline{k}\overline{a})+\left(cos(\overline{k}\overline{a})+\frac{2\mu \overline{V}_{o}}{\hbar^2\overline{k}}sin(\overline{k}\overline{a})\right)^2
\end{equation}
this value of A$^2$(k) can be interpreted as the ratio of probability of finding the particle for a particular value of k. Taking the superposition of scattering states with $\overline{k}$ near $\overline{k_{n}}$ using scattering state method and expanding A$^2$ about $\overline{E}_{n}$ = $\frac{\hbar^2\overline{k_{n}^2}}{2\mu}$ we can write as
\begin{eqnarray}
A^2(\overline{E})\approx \left(\frac{\mu \overline{a}}{\hbar^2 \overline{k_{n}}}\right)^2\left[1+\left(\frac{2 \mu \overline{V}_{o}}{\hbar^2 \overline{k_{n}}}\right)^2\right](\overline{E}-\overline{E}_{n}+\delta)^2+\left[1+\left(\frac{2 \mu \overline{V}_{o}}{\hbar^2 \overline{k_{n}}}\right)^2\right]^{-1}\nonumber \\
\equiv D^2(\Delta+\delta)^2+H^2
\end{eqnarray}
 where
\begin{eqnarray}
\Delta = \overline{E}-\overline{E}_{n}\nonumber \\
and \nonumber \\
D^2 = \left(\frac{\mu \overline{a}}{\hbar^2 \overline{k_{n}}}\right)^2\left[1+\left(\frac{2 \mu \overline{V}_{o}}{\hbar^2 \overline{k_{n}}}\right)^2\right] \nonumber \\
H^2 = \left[1+\left(\frac{2 \mu \overline{V}_{o}}{\hbar^2 \overline{k_{n}}}\right)^2\right]^{-1} \nonumber \\
\delta = \frac{2 \overline{V}_{o}}{\overline{a}}\left[1+\left(\frac{2 \mu \overline{V}_{o}}{\hbar^2 \overline{k_{n}}}\right)^2\right]^{-1}
\end{eqnarray}
now the scattering states having $\overline{E}$ near $\overline{E}_{n}$ can be written as
\begin{align*}
\begin{cases}
\phi_{\Delta}(\overline{x}) = \sqrt{\frac{2}{J}}\frac{1}{\sqrt{D^2(\Delta+\delta)^2+H^2}}sin(\overline{k}\overline{x}),\; \overline{x}<\overline{a}\\
\phi_{\Delta}(\overline{x}) = cos(\overline{k}\overline{x}+\theta),\; \overline{a}<\overline{x}
\end{cases}
\end{align*}
this system is quantized between [0, J], where J$>\overline{a}$. From these scattering states an initial state is constructed in such a way that
\begin{gather*}\begin{cases} 
\psi(\overline{x},\tau = 0) = \sum_{\Delta} c_{\Delta}\phi_{\Delta}(\overline{x})=\psi_{n}(\overline{x}),\; \overline{x}<\overline{a}\\
\psi(\overline{x},\tau = 0) = \sum_{\Delta} c_{\Delta}\phi_{\Delta}(\overline{x})=0,\; \overline{x}>\overline{a}
\end{cases}\end{gather*}
coefficient $c_{\Delta}$ can be calculated from the orthogonality of states $\phi_{\Delta}(\overline{x})$,
\begin{eqnarray}
c_{\Delta} = \int_{o}^{J}d\overline{x}\phi_{\Delta}(\overline{x})\psi(\overline{x},0) \nonumber \\
= \sqrt{\frac{2}{J}}\frac{1}{\sqrt{D^2(\Delta+\delta)^2+H^2}}\int_{0}^{\overline{a}}d\overline{x}sin(\overline{k}\overline{x})\psi_{n}(\overline{x})
\end{eqnarray}
 If we choose 
\begin{equation}
\psi_{n}(\overline{x})\approx \sqrt{\frac{2}{\overline{a}}}sin\left(\frac{n\pi \overline{x}}{\overline{a}}\right), n = 1, 2,3..
\end{equation}
we get 
\begin{eqnarray}
c_{\Delta} \approx \frac{1}{\sqrt{J \overline{a}}}\frac{1}{\sqrt{D^2(\Delta+\delta)^2+H^2}}\int_{0}^{\overline{a}}d\overline{x}sin(\overline{k}\overline{x})sin\left(\frac{n\pi \overline{x}}{\overline{a}}\right)\nonumber \\
\approx \sqrt{\frac{\overline{a}}{J}}\frac{1}{\sqrt{D^2(\Delta+\delta)^2+H^2}}
\end{eqnarray}
hence initial state is then given by 
\begin{equation}
\psi(\overline{x},\tau = 0) \approx \sqrt{{\frac{\overline{a}}{J}}}\sum_{\Delta}\frac{1}{\sqrt{D^2(\Delta+\delta)^2+H^2}}\phi_{\Delta}(\overline{x})
\end{equation}
from equation $(13)$,$(15)$ and $(16)$, the solution at a later time $\tau$ is given by 
\begin{equation}
\psi(\overline{x},\tau) \approx \sqrt{{\frac{\overline{a}}{J}}}\sum_{\Delta}\frac{1}{\sqrt{D^2(\Delta+\delta)^2+H^2}}\phi_{\Delta}(\overline{x})e^{\left(\frac{i}{\hbar}\right)\left(\overline{E}_{n}+\Delta\right) \tau}
\end{equation}
since the system is quantized in the interval [0, J], we have $\overline{k}J = n^{'}\frac{\pi}{2}$. Replacing the sum by an integral 
\begin{equation}
\sum_{\Delta}\rightarrow \int d\Delta \frac{J}{\pi \hbar}\sqrt{\frac{2\mu}{\overline{E}_{n}}}
\end{equation}
equation $(36)$ can be written as
\begin{equation}
\psi(\overline{x},\tau) \approx \int_{\-\infty}^{\infty} d\Delta \frac{J}{\pi \hbar}\sqrt{\frac{2\mu}{\overline{E}_{n}}} \sqrt{{\frac{\overline{a}}{J}}}\frac{1}{\sqrt{D^2(\Delta+\delta)^2+H^2}}\phi_{\Delta}(\overline{x})e^{\left(\frac{i}{\hbar}\right)\left(\overline{E}_{n}+\Delta\right) \tau}
\end{equation}
using the value of $\phi_{\Delta}(\overline{x})$from the preceding expressions we can approximate the wavefunction as
\begin{eqnarray}
\psi(\overline{x},\tau) \approx \frac{2}{\pi \hbar}\sqrt{\frac{\mu \overline{a}}{\overline{E}_{n}}}sin(\overline{k_{n}}\overline{x})e^{-(\frac{i}{\hbar})\overline{E}_{n}\tau} \times \int_{\-\infty}^{\infty} d\Delta\frac{e^{-(\frac{i}{\hbar})\Delta \tau}}{\sqrt{D^2(\Delta+\delta)^2+H^2}} \nonumber \\ = \frac{2}{\hbar}\sqrt{\frac{\mu \overline{a}}{\overline{E}_{n}}}\frac{sin(\overline{k_{n}}\overline{x})}{|HD|}e^{-\left(\frac{i}{\hbar}\right)\overline{E}_{n}\tau}e^{-\left(\frac{1}{\hbar}\right)|\frac{H}{D}|\tau}
\end{eqnarray}
Now the probability of finding the particle is given by 
\begin{equation}
P(t) = \frac{\int_{0}^{a(t)}|\phi(x,t)|^2dx}{\int_{0}^{a(0)}|\phi(x,0)|^2dx} = \frac{\int_{0}^{\overline{a}}|\psi(\overline{x},\tau)|^2d\overline{x}}{\int_{0}^{\overline{a}}|\psi(\overline{x}, 0)|^2d\overline{x}}
\end{equation}
 from equation $(39)$ we find that the probability is 
\begin{equation}
P(t)\approx exp \left[-\alpha_{n}(t)\right]
\end{equation}
and non adiabatic transition probability is given by
\begin{equation}
P_{(non-adiabatic)} = 1- P(t) = 1- exp \left[-\alpha_{n}(t)\right]
\end{equation}
where 
\begin{equation}
\alpha(t) = 2\left|\frac{H}{D}\right|\frac{t}{R_{o}R(t)}
\end{equation}
using equation $(10)$ and $(31)$ above equation can be further written as
\begin{equation}
\alpha(t) = 2\left(\frac{\hbar^2 \overline{k_{n}}}{\mu \overline{a}}\right)\left[1+\frac{2\mu \overline{V}_{o}}{\hbar^2\overline{k_{n}}}\right]^{-1}\frac{t}{R_{o}R(t)}
\end{equation}
\section{Conclusions}
Considering the time-dependent scaling factor as reported by Berry and klein we consider the crossing of two diabatic potentials with coupling as a moving $\delta$ potential. Using this factor we have transformed our time dependent Schrodinger equations into time independent one which are further reduced into single equation which contains a Green's function having the effect of second state in it.
This equation is further solved by the reported methods to find out the expressions for non adiabatic transition probability. To conclude the problem of two state non adiabatic transition probability with a moving $\delta$ potential coupling can be handled by the time scaling factor.

\section{Acknowledgments:}
\noindent The authors thank Indian Institute of Technology Mandi for providing Professional development fund along with HTRA scholarship.

\end{document}